\documentclass[10pt,twocolumn,floatfix]{revtex4-1} 
\usepackage{verbatim}
\usepackage{graphicx}
\usepackage{sidecap}
\usepackage{array}
\usepackage{multirow}
\usepackage{subfigure}
\usepackage{color}

\begin{document}
\title{Effects of finite coverage on global polarization observables in heavy ion collisions}
\author{
Shaowei Lan$^{1}$,
Zi-Wei Lin$^{1,2}$,
Shusu Shi$^{1}$,
Xu Sun$^{1}$
}
\affiliation{$^{1}$Key Laboratory of Quarks and Lepton Physics (MOE) and Institute of Particle Physics, Central China Normal University, Wuhan, 430079, China}
\affiliation{$^{2}$Department of Physics, East Carolina University, Greenville, North Carolina 27858, USA}
%\linenumbers

\begin{abstract}
In non-central relativistic heavy ion collisions, the created matter 
possesses a large initial orbital angular momentum. 
Particles produced in the collisions could be polarized globally in the direction of the orbital angular momentum due to spin-orbit coupling.
Recently, the STAR experiment has presented polarization signals
for  $\Lambda$ hyperons and possible spin alignment signals for
$\phi$ mesons. 
Here we discuss the effects of finite coverage on these observables. 
The results from a multi-phase transport and a toy model both indicate that 
a pseudorapidity coverage narrower than $|\eta|< \sim 1$ 
will generate a larger value for the extracted $\phi$-meson
$\rho_{00}$ parameter; thus a finite coverage can lead to an
artificial deviation of $\rho_{00}$ from 1/3. 
We also show that a finite $\eta$ and $p_T$ coverage affect the
extracted $p_H$ parameter for $\Lambda$ hyperons 
when the real $p_H$ value is non-zero.
Therefore proper corrections are necessary to reliably
quantify the global polarization with experimental observables. 
\end{abstract}

\maketitle

%\section{Introduction}
{\em Introduction.}
The dense matter created in non-central relativistic heavy ion
collisions processes a large orbital angular momentum. 
It has been predicted that the orbital angular moment may result in
a net-polarization of produced particles along the direction of the
initial angular moment due to the spin-orbit
coupling~\cite{Liang_Wang1,Liang_Wang2}.  Recently, the RHIC-STAR
collaboration claimed the discovery of a global $\Lambda$ polarization
in heavy ion collisions~\cite{STAR_LamP}. 
The estimated vorticity is much larger than all other known fluids,
which suggests that the quark-gluon plasma is the most vortical fluid
produced in the laboratory.  Meanwhile the measured spin alignment
parameter $\rho_{00}$ of $\phi$ mesons is systematically larger than
1/3~\cite{STAR_phispin1, STAR_phispin2}. 
A deviation of $\rho_{00}$ from 1/3 indicates a spin alignment of the
vector mesons, and whether $\rho_{00}$ is greater or smaller than 1/3
depends on the hadronization mechanism~\cite{Liang_Wang2}.  
More detailed measurements of polarization observables, e.g., 
as a function of rapidity, pseudorapidity ($\eta$), or transverse momentum ($p_T$),
will be important to better understand the novel polarization phenomena.

The angular distribution of the $\Lambda$ decay products with respect
to the orbital angular momentum of system $\textbf{L}$ 
can be written as ~\cite{Lam_pH}
\begin{equation}
\frac{dN}{d\cos\theta^{\star}} \propto 1 + \alpha_{H}p_H\cos\theta^{\star},
\label{Equation:pH0}
\end{equation}
where $\theta^{\star}$ is the angle between $\textbf{L}$ and the
momentum of the daughter proton in the rest frame of the parent
$\Lambda$ hyperon, and $\alpha_{H}$ is the decay parameter 
$\alpha_{\Lambda}$ = -$\alpha_{\Lambda}$ = 0.642 $\pm$ 0.013~\cite{pdg2014}.
Therefore the global polarization parameter $p_H$ for $\Lambda$ can be calculated by
\begin{equation}
p_H = \frac{3}{\alpha_{H}}\left \langle \cos \theta^{\star} \right \rangle.
\label{Equation:pH1}
\end{equation}
The angle brackets above represent the averaging over all $\Lambda$ decays. 
Assuming a perfect detector acceptance, one can also write~\cite{STAR_Lam}
\begin{equation}
p_H = \frac{8}{\pi \alpha_{H}}  \left \langle \cos (\phi^{\star}_{p} -
  \phi_{\textbf{L}}) \right \rangle, 
\label{Equation:pH2}
\end{equation}
where $\phi^{\star}_{p}$ is the azimuth of the daughter proton
momentum vector in the $\Lambda$ rest frame and $\phi_{\textbf{L}}$ is
the azimuth of the system angular momentum. 

For vector mesons such as the $\phi$ meson, the polarization
information including the spin alignment is described by a spin density matrix $\rho$. 
A deviation of the diagonal element $\rho_{m,m}$ ($m = -1, 0, 1$) from
1/3 is the signal of spin alignment. 
For a diagonal matrix $\rho$, since the elements $\rho_{-1,-1}$ and
$\rho_{1,1}$ are degenerate,
$\rho_{0,0}$ is the only independent element~\cite{STAR_phi}. 
It can be extracted from the following angular distribution of the decay
products of vector mesons with respect to $\textbf{L}$~\cite{phi_rho00}: 
\begin{equation}
\frac{dN}{d\cos\theta^{\star}} \propto (1-\rho_{00}) + (3\rho_{00}-1)\cos^{2}\theta^{\star}.
\label{Equation:rho00}
\end{equation}

So far lots of theoretical progresses have been made on the
evolution of vorticity and polarization~\cite{theory1, theory2,
  theory3, theory4}. 
Since the experimental coverage of the phase space is always finite, 
in this paper we study  the effects of a finite $\eta$ and $p_T$ coverage
on the measured global polarization parameter $p_H$ of $\Lambda$ hyperons
and spin alignment parameter $\rho_{00}$ of $\phi$ mesons. 

{\em The simulation method.}
We have modified a multi-phase transport
(AMPT) model~\cite{AMPT} for this study. 
The string melting version of the AMPT model~\cite{smAMPT} used here
consists of a fluctuating initial condition, an elastic 
parton cascade, a quark coalescence model for hadronization, and a
hadron cascade. 
We use the same model parameters as those used for the top RHIC energy
in an earlier study~\cite{amptparam}.
In the hadron cascade of AMPT, $\phi$ mesons and $\Lambda$ hyperons
are assumed to be uniformly polarized according to Eqs.(\ref{Equation:pH0}) and
(\ref{Equation:rho00}) when they decay. 
The included channels of polarized decays 
are $\phi \to K+\bar{K}$ (branching ratio
$\sim$ 83\%) and $\Lambda \to p+\pi^{-}$ and  $n+\pi^{0}$ (branching ratio
$\sim$ 100\%).  While experimentally the direction of
the angular momentum is often estimated by the normal of the
reconstructed event plane~\cite{STAR_Lam, STAR_phi}, 
we directly calculate the initial angular momentum vector of 
the participant nucleons event-by-event in the AMPT calculations. 
All results in this study are for minimum bias 
(impact parameter from $0$ to $15.6$ fm) Au+Au collisions.

\begin{figure}[]
\vskip 0.0cm
\centering
\centerline{\includegraphics[scale=0.45]{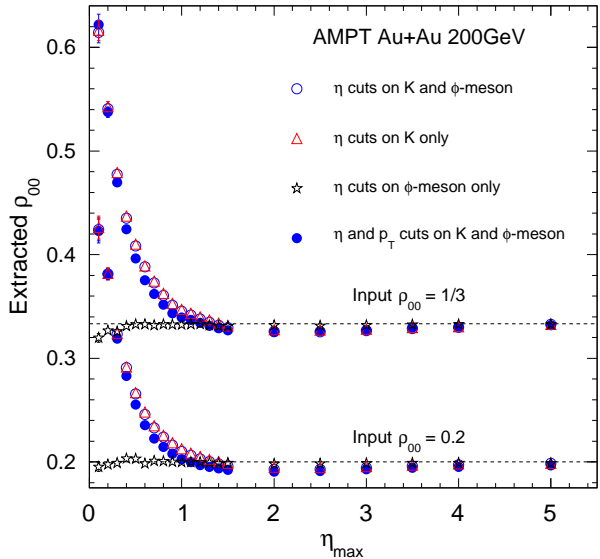}}
\caption{(Color online) The extracted $\rho_{00}$ as a function of the
  upper limit of the $|\eta|$ coverage from AMPT for minimum bias
  Au+Au collisions at $\sqrt{s_{NN}}$ = 200 GeV, 
where dashed lines represent the input $\rho_{00}$ value of 0.2 or
1/3. Results for $\eta$ cuts applied to both kaons and $\phi$ mesons, only kaons, 
  and only $\phi$-mesons are shown, in addition to results with
  both the $\eta$ cuts and the STAR $p_T$ cuts on kaons and
  $\phi$-mesons.
}
\label{fig1}       
\end{figure}

\begin{figure*}[ht]
\vskip 0.0cm
\centering
\centerline{\includegraphics[scale=0.8]{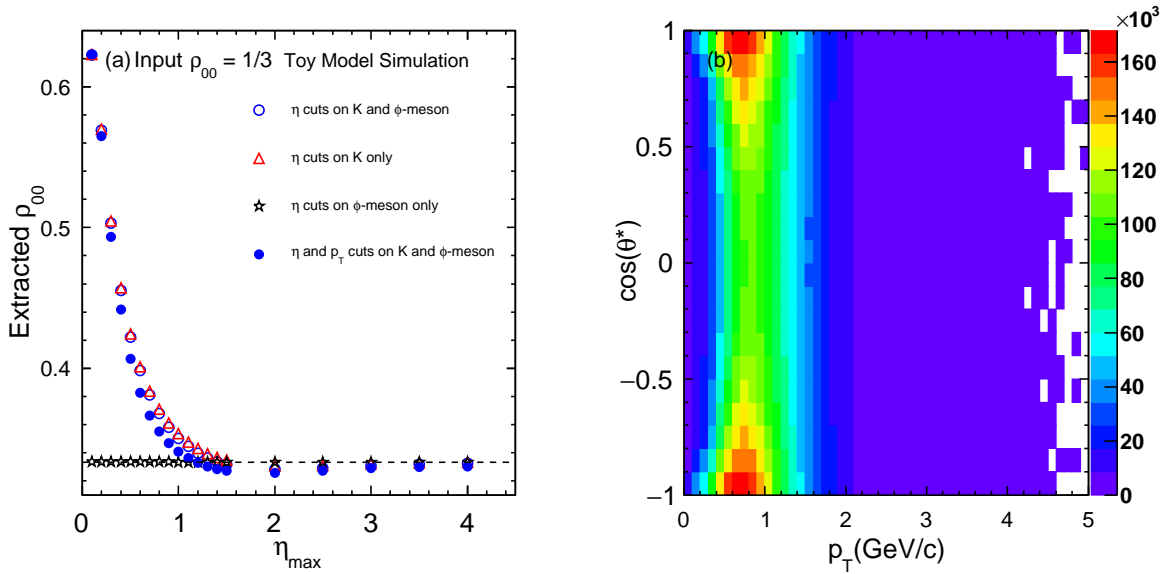}}
\caption{(Color online) Effects of $\eta$ and $p_T$ cuts on $\rho_{00}$ for
  $\phi$-meson decays from the toy model.
(a) The extracted $\rho_{00}$ as a function of $\eta_{\rm max}$. 
(b) Distribution in $\cos\theta^{\star}$ and $p_T$
for kaons from $\phi$ decays after applying the cut $|\eta| <
0.3$ to kaons.
}
\label{fig2}    
\end{figure*}   

\begin{figure}[]
\vskip 0.0cm
\centering
\centerline{\includegraphics[scale=0.45]{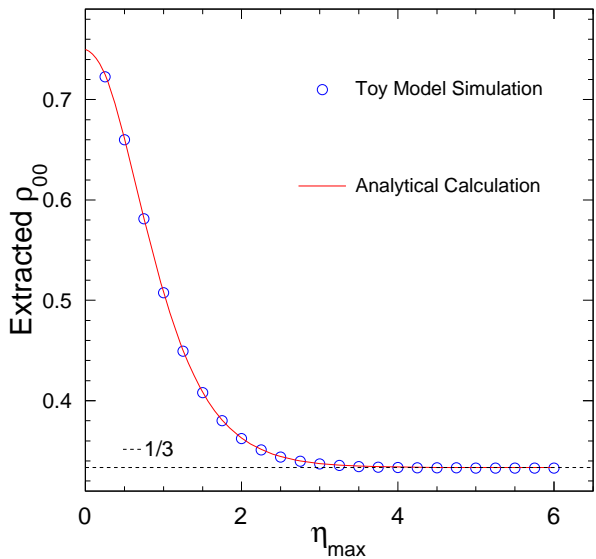}}
\caption{(Color online) 
The extracted $\rho_{00}$ as a function of $\eta_{\rm max}$ 
for $\phi$ mesons at rest from the toy model (circles) and 
analytical calculations (solid line).
}
\label{fig3}    
\end{figure}

\begin{figure*}[ht]
\vskip 0.0cm
\centering
\centerline{\includegraphics[scale=0.9]{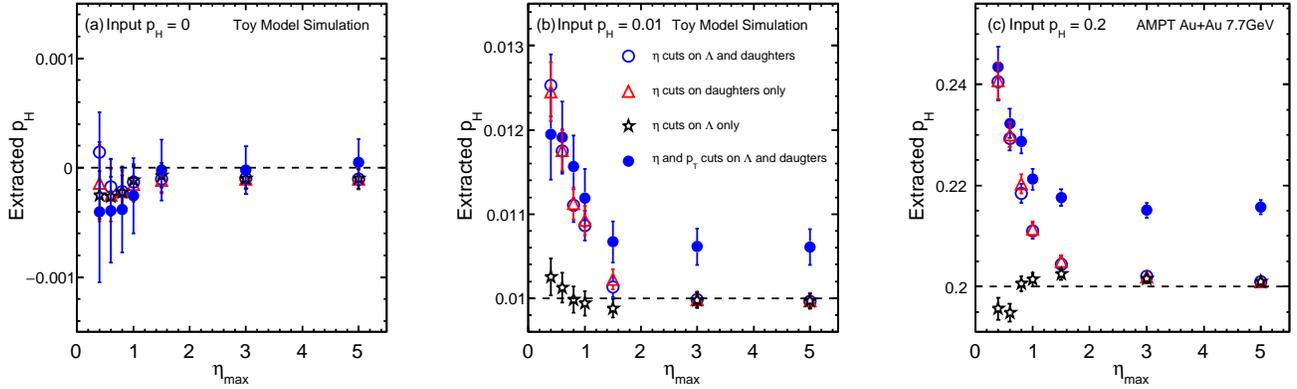}}
\caption{(Color online) The $p_H$ parameter extracted with 
Eq.(\ref{Equation:pH2}) 
as a function of $\eta_{\rm max}$ 
for Au+Au collisions at $\sqrt{s_{NN}}$=7.7 GeV 
from (a) the toy model with input value of 0,
(b) the toy model with input value of 0.01,
and (c) AMPT with input value of 0.2.
Filled circles represent the results with both $\eta$ cuts 
and the STAR $p_T$ cuts.
}
\label{fig4}    
\end{figure*}  

\begin{figure}[]
\vskip 0.0cm
\centering
\centerline{\includegraphics[scale=0.45]{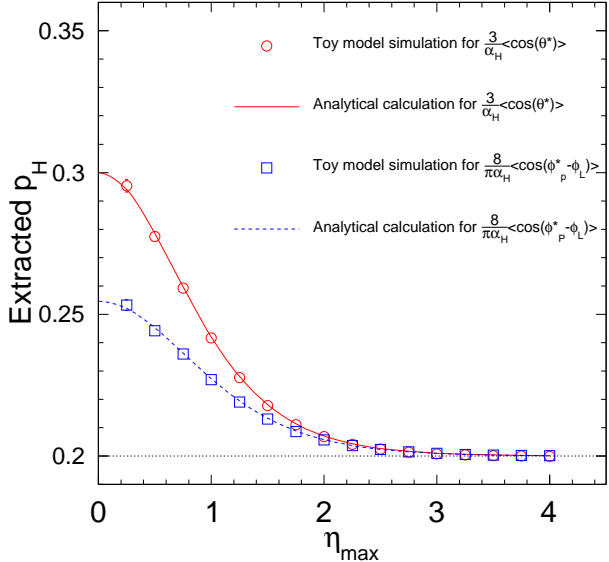}}
\caption{(Color online) 
The $p_H$ extracted with Eq.(\ref{Equation:pH1}) (circles) or
Eq.(\ref{Equation:pH2}) (squares) as a function of $\eta_{\rm max}$ 
from the toy model for $\Lambda$ hyperons at rest, in comparison with
the analytical results 
(curves), for the input value $p_H=0.01$ (dotted line).  
}
\label{fig5}    
\end{figure}

\begin{figure*}[ht]
\vskip 0.0cm
\centering
\centerline{\includegraphics[scale=0.8]{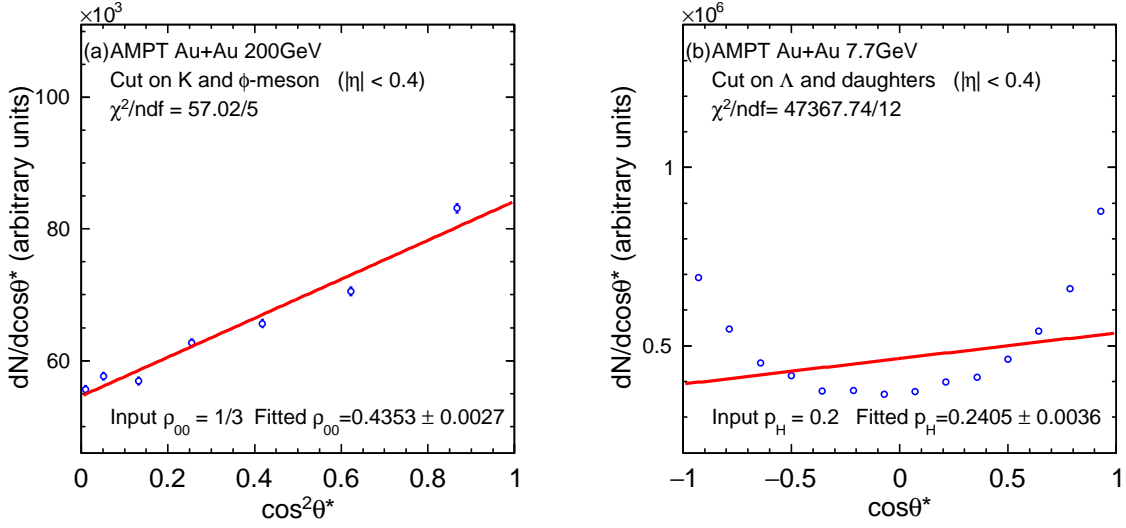}}
\caption{(Color online) 
Examples of the effect of the cut $|\eta|< 0.4$ on the shape of 
(a) the $\cos^{2}\theta^{\star}$ distribution for $\phi$ mesons,
and (b) the $\cos\theta^{\star}$ distribution for $\Lambda$ hyperons. 
}
\label{fig6}    
\end{figure*}

\begin{figure}[]
\vskip 0.0cm
\centering
\centerline{\includegraphics[scale=0.45]{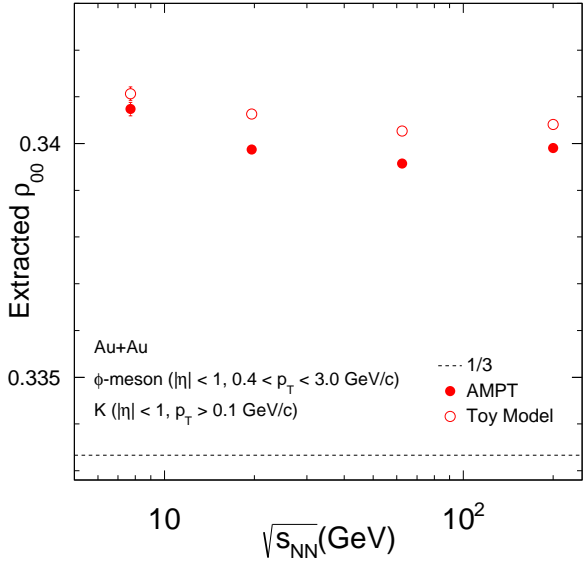}}
\caption{(Color online) The extracted $\rho_{00}$ of $\phi$ mesons
  from AMPT and the toy model for the input value $\rho_{00}=1/3$ for
  Au+Au collisions at  different energies.
}
\label{fig7}    
\end{figure}

\begin{figure}[]
\vskip 0.0cm
\centering
\centerline{\includegraphics[scale=0.45]{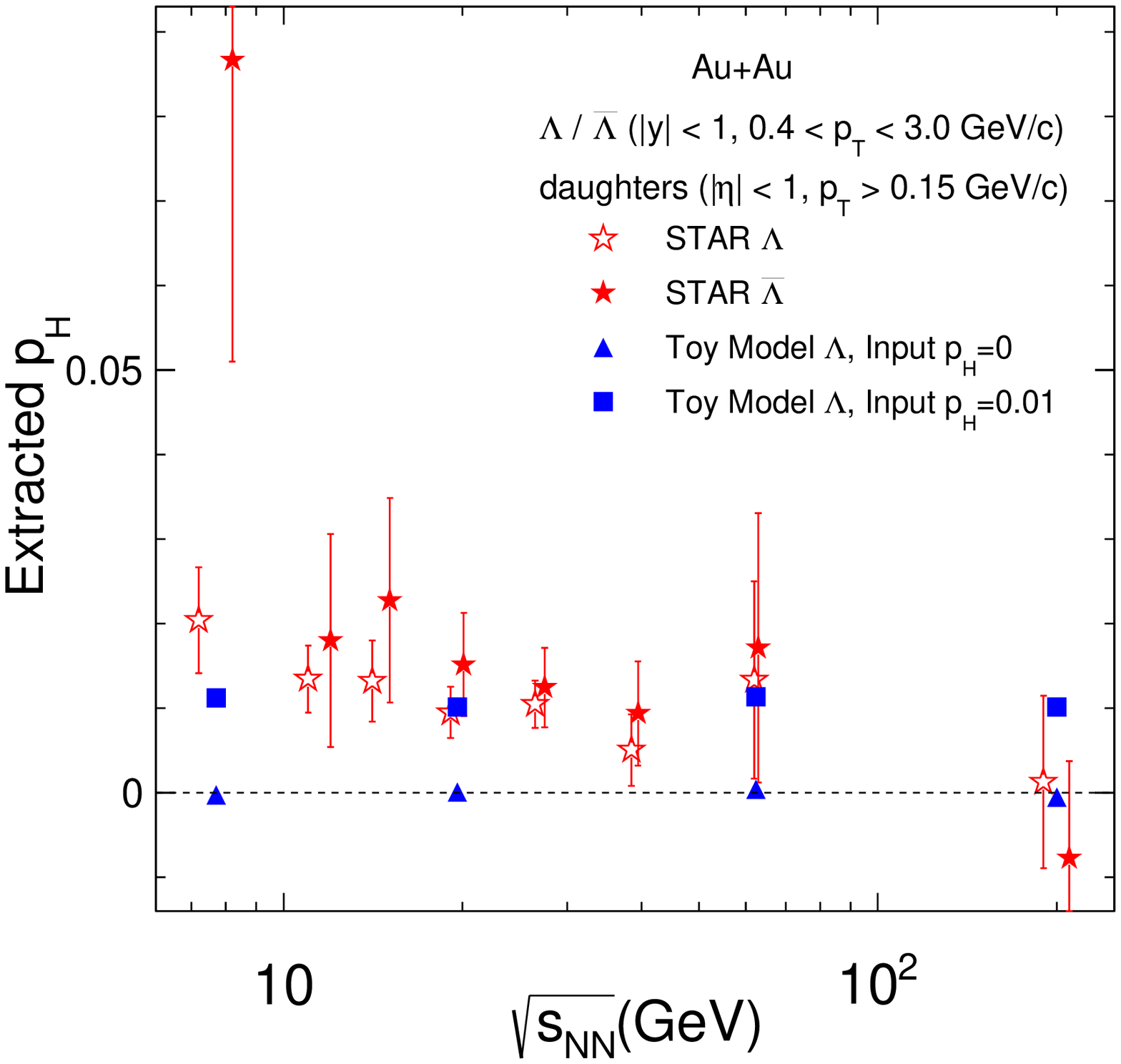}}
\caption{(Color online) The extracted $p_H$ of $\Lambda$ hyperons 
  from the toy model for the input value $p_H=0$ and $p_H=0.01$ 
 for Au+Au collisions at  different energies in comparison with the
 STAR data.
}
\label{fig8}    
\end{figure}

We follow the experimental procedure to extract the $\rho_{00}$ parameter
of $\phi$ mesons: pairs of $K\bar{K}$ from
each event are used to reconstruct the $\phi$-meson candidates, while
pairs from different events are constructed to estimate the
background; these same event pairs and mixed event pairs are both
divided into multiple $\cos\theta^{\star}$ bins.
We then extract the $\rho_{00}$ parameter by using
Eq.~(\ref{Equation:rho00}) to fit the $|\cos\theta^{\star}|$
distribution of the $\phi$-meson signal, i.e., the difference between the
same event and mixed event pairs. 
For $\Lambda$ hyperons, 
we found that the systematic uncertainty of background subtraction 
(up to a fraction of a percent) is too large when the input $p_H$ is
small or zero; it is also not straightforward to apply 
topological cuts to decays from the AMPT model as done in the experimental analysis. 
Therefore we use the $\Lambda$ decay information from the AMPT model directly 
to extract the $p_H$ parameter. 
We have found that hadronic rescatterings have negligible effects on
the value of the extracted $p_H$ parameter for $\Lambda$ hyperons or
the $\rho_{00}$ parameter for $\phi$ mesons; this is
expected due to their long lifetimes before decay. 
We have also used the $\phi$ decay information directly 
to extract the $\rho_{00}$ parameter from the AMPT model and obtained
consistent values as those extracted from the background subtraction method.

When there are no phase space cuts on particles (i.e. when we have
full coverage), we have found that the extracted $\rho_{00}$ or $p_H$
value is consistent with the input value as expected. 
However, the experimental acceptance is always limited. 
The detected $\eta$ range for finial state particles is typically from
$\pm 0.5$ to $\pm 2$, 
e.g., the STAR experiment covers $|\eta| < 1$; and the $p_T$ of
detected particle tracks is typically larger than 0.1 - 0.2 GeV/$c$. 
Not only are the candidate decay daughters for $\phi$ and $\Lambda$
reconstructions within a certain $\eta$ and $p_T$ coverage, 
but additional $y$ (or $\eta$) and $p_T$ cuts are also often applied to the
reconstructed parent particles ($\phi$ and $\Lambda$). 
Therefore we need to investigate whether and how the phase-space cuts may
influence the extracted value of the global polarization/spin-alignment
parameters. Note that the vorticity in a nuclear collision depends on the
transverse position and pseudorapidity in principle\cite{theory4}, thus the real
polarization may depend on the phase space variables such as
$\eta$ and $p_T$; but that is outside the scope of this study.

{\em Results.}
Figure~1 shows the AMPT model results for the extracted $\rho_{00}$ of
$\phi$ mesons as a function of the upper limit of the $|\eta|$
coverage, $\eta_{\rm max}$, for Au+Au collisions at $\sqrt{s_{NN}}$ = 200 GeV. 
Results are shown for two input $\rho_{00}$ values, 0.2 and 1/3,
where we study the $\eta$-cut effects by applying the cut $|\eta|<\eta_{\rm max}$ 
to both kaons and $\phi$ mesons (open circles), only kaons
(triangles), or only $\phi$ mesons (stars). 
We see that, 
when the $|\eta|$ cut is only applied to the parent $\phi$ mesons, the
extracted $\rho_{00}$ has almost no deviation from the input value for
any $\eta$ cut. 
On the other hand, the extracted $\rho_{00}$ strongly depends on the
$\eta$ cut applied to the decay daughter candidates, where 
an $\eta$ coverage narrower than $\sim 1$ gives a significantly
larger extracted $\rho_{00}$ than the input value. 
This is because, with the angular momentum $\textbf{L}$ in the
transverse plane, a narrow $\eta$ cut on kaons tends to exclude some kaons
along the beam directions and thus excludes those $\phi$-meson
decays with daughter kaons around $\theta^{\star} \sim 90^{\circ}$.
Note that such loss of decay daughters close to the beam direction due to finite acceptance 
and its effect on the hyperon global polarization parameter $p_H$ have
been pointed out earlier for Lambda decays~\cite{STAR_Lam}.
Also shown in Fig. 1 are the results where both the $\eta$ cuts
and the STAR $p_T$ cuts ($p_T^K >$ 0.1 GeV/$c$ and 0.4
$<p_T^{\phi}<$ 3 GeV/$c$) are applied to the kaons and $\phi$ mesons
(filled circles), 
where we see that the $p_T$ cuts lead to a small reduction of the
extracted $\rho_{00}$ values. We also see in Fig.1 that the extracted
$\rho_{00}$ value converges to the input value at large $\eta_{\rm max}$ for
all the considered cuts. 

To further illustrate the $\eta$-cut effect on $\rho_{00}$, 
we also use a toy model, where we sample the $p_T$, $\eta$ and azimuth
distributions of $\phi$ mesons according to the AMPT results and decay
the hadrons with PYTHIA~\cite{PYTHIA}.  
As shown in Fig.2a, the effects of $\eta$ cuts from the toy model are
essentially the same as those from the AMPT model. 
Figure~2b shows the two-dimensional distribution 
(in $\cos\theta^{\star}$ and $p_T$) of kaons within $|\eta| < 0.3$
from $\phi$ decays. We can clearly observe that the $\eta$ cut excludes more kaons from
$\phi$ decays around $\theta^{\star} \sim 90^{\circ}$,
i.e., around $\cos\theta^{\star} \sim 0$. 
The figure also demonstrates that the $\eta$-cut mostly affects the low $p_T$ region. 
Furthermore, a different $\eta$-cut effect is observed for
$\eta_{max}$ 
roughly between 1.3 and 4 from both AMPT and the toy model, where
the extracted $\rho_{00}$ can be smaller than the input value. 
This could be the result of the finite $p_T$ and finite anisotropic
flow of the parent $\phi$ mesons, which complicate the relation between the
$\eta$ cut (in the center-of-mass frame) and the $\theta^{\star}$
distribution of the decay daughters (in the $\phi$ rest frame). 
As a test, we show in Fig.3 the effect of $\eta$ cuts on $\phi$
mesons at rest from the toy model (circles), where the extracted
$\rho_{00}$ monotonically decreases with $\eta_{\rm max}$ and finally
approaches the input value (1/3 here). 
The solid line in Fig.3 represents the analytical result on the
$\eta$-cut effect for $\phi$ mesons at rest, as shown in the appendix,
which agrees with the toy model result. 

For $\Lambda$ hyperons, Fig.4 shows the $\eta$-cut effect on the $p_H$
parameter when Eq.(\ref{Equation:pH2}) is used to extract $p_H$,
as often done in the experimental
measurements~\cite{STAR_LamP,STAR_Lam}.
Figures 4a and 4b show the toy model results for the input value
$p_H=0$ and  $p_H=0.01$, respectively; while  
Fig.4c shows the AMPT results for the input value $p_H=0.2$. 
In the toy model, we sampled $\Lambda$ hyperons using 
$p_T$, $\eta$ and azimuth distributions that have been fitted to the
AMPT results, where the azimuth distribution includes the elliptic flow.
Note that the case $p_H$ = 0 represents no real global polarization, 
while the case $p_H$=0.01 roughly corresponds to the STAR data~\cite{STAR_LamP}. 
When the input $p_H$ is zero, 
we see that the extracted $p_H$ values are consistent with zero, 
with or without the phase space cuts. 
This is in line with the expectation from
the up-down symmetry with respect to the direction of $\bf{L}$. 
However, when the input $p_H$ is finite (and positive), 
the effects of $\eta$ cuts on $p_H$ for $\Lambda$ hyperons
are similar as those on $\rho_{00}$ for $\phi$ mesons:
a finite $|\eta|$ cut applied to the decay-daughter candidates
generates an extracted $p_H$ value larger than the input value, 
but a finite $|\eta|$ cut applied to parent $\Lambda$ hyperons has
little effect on the extracted $p_H$.
Also, the extracted $p_H$ value approaches the input value 
for $\eta_{\rm max}$ greater than about 2 (when there are no $p_T$ cuts).

We also see from Figs.4b and 4c that $p_T$ cuts can lead to 
deviations of the extracted $p_H$ from the input value,   
even when there is no $\eta$ cut. This is quite different from 
the $\phi$-meson case, where the extracted $\rho_{00}$ value 
is not affected by the $p_T$ cuts when there is no $\eta$ cut.
Note that the $p_T$ cuts in Fig.4 are the STAR cuts: $p_T^{\rm daughter} >
0.15$ GeV/$c$ and $0.4 <p_T^{\Lambda}<3$ GeV/$c$. 
Furthermore, we see that the effects of the phase-space cuts from the 
AMPT model for input $p_H= 0.2$ in Fig.4c are qualitatively the same 
as those from the toy model for input $p_H=0.01$ in Fig.4b.

We have analytically derived the $\eta$-cut effect on the extracted
$p_H$ value for $\Lambda$ decays at rest (see the appendix), 
which is shown as lines in Fig.5 for the input value $p_H=0.01$ when 
the $p_H$ value is calculated with Eq.(\ref{Equation:pH1}) (solid) or
Eq.(\ref{Equation:pH2}) (dashed). 
Symbols represent the corresponding results from the toy model, which
agree with the analytical results. 
We also see that the $\eta$-cut effect is smaller when
using Eq.(\ref{Equation:pH2}) instead of Eq.(\ref{Equation:pH1}). 

{\em Discussions and summary. }
While the global spin alignment and polarization signals can be
respectively described by Eq.(\ref{Equation:pH0}) and
Eq.(\ref{Equation:rho00}), 
phase-space cuts such as the $\eta$ and $p_T$ cuts discussed here 
modify the functional forms.
We show in Fig.6 examples after the cut $|\eta| < 0.4$ is applied, 
where the $\cos\theta^{\star}$ distribution for $\phi$ mesons (when
plotted versus $\cos^2\theta^{\star}$) and the $\cos\theta^{\star}$
distribution for $\Lambda$ hyperons deviate from a straight line with
a large $\chi^2$ per degree of freedom.  
This method can be used in the experimental analysis to probe the effect of
phase-space cuts, e.g., by selecting a narrow $\eta$ range within the
experimental coverage and observing a deviation from the expected straight line.

A natural question is whether the experimentally observed signals of spin
alignment and polarization could be mostly due to the phase-space
cuts. 
Figure~7 shows the results on the extracted $\rho_{00}$ for $\phi$
mesons  from AMPT and the toy model for the case of no spin
alignment (i.e. input $\rho_{00}=1/3$) after the STAR $\eta$ and $p_T$
cuts are applied.
We see that the extracted $\rho_{00}$ parameters are
systematically higher than 1/3 and have a weak energy dependence,
similar to the STAR preliminary measurements~\cite{STAR_phispin1, STAR_phispin2}. 
This indicates that the effect of the phase-space cuts may be a 
dominant contribution to the deviation of the current experimental
data of $\rho_{00}$ from 1/3.

For $\Lambda$ hyperons, we show in Fig.8 the toy model results from 
Eq.(\ref{Equation:pH2}) for the input value $p_H=0$ (triangles) and
$p_H=0.01$ (squares),  
where the STAR phase-space cuts have been applied.
We see that the extracted $p_H$ values are consistent with zero when 
there is no global polarization (for input $p_H$=0). 
With input $p_H$ = 0.01, the extracted $p_H$ values are rather close to the input
value (within $\sim$ 0.1\%), 
and the phase-space cuts do not introduce an 
energy dependence for the extracted $p_H$ values.
The STAR data~\cite{STAR_LamP} on the $p_H$ values of $\Lambda$ and
$\bar \Lambda$ are shown for comparisons, where the effects of 
phase-space cuts have already been corrected~\cite{STAR_LamP, STAR_Lam}.

In summary, we have used a modified AMPT model that includes the
decays of polarized $\phi$ mesons and $\Lambda$ hyperons as well as a
toy model to study the effects of phase-space cuts on the extracted
polarization parameters $\rho_{00}$ and $p_H$.  
We find that a finite $\eta$ coverage narrower than $\eta < \sim 1$
leads to a larger extracted value of $\rho_{00}$ than the input value,
and a similar behavior is observed for the extracted $p_H$ when the
input $p_H$ is positive. The narrower the $\eta$
coverage, the larger the extracted values. 
A finite $p_T$ coverage also affects the extracted values of
$\rho_{00}$ and $p_H$.
When we assume no global polarization by setting the input 
$\rho_{00}$ to 1/3, the model results after the STAR $\eta$ and $p_T$
cuts give extracted $\rho_{00}$ values of $\phi$ mesons that significantly
deviate from the value 1/3.  
These extracted $\rho_{00}$ values are similar to the preliminary
experimental data, suggesting that the phase-space cuts may be a
dominant contribution to the observed higher-than-1/3 values. 
For $\Lambda$ hyperons, we find that, when assuming no polarization,
a finite coverage in $\eta$ and/or $p_T$ does not lead to a non-zero
extracted $p_H$; however $\eta$ and $p_T$ cuts affect the magnitudes
of the $\Lambda$ polarization parameter $p_H$ when the input $p_H$ is
non-zero. Our study thus shows that measured polarization observables need to be
corrected for the effects of finite acceptance before quantitative
conclusions on the global polarization can be reliably made.

%\section*{Acknowledgments}
{\em Acknowledgments}
We thank Mike Lisa and Nu Xu for discussions. This work was supported in part by the National Basic Research Program of China (973 program) under Grant No. 2015CB8569, 
National Natural Science Foundation of China under Grants No. 11628508 and No. 11475070 and China Postdoctoral Science Foundation under Grants No. 2016M592357.

%%%%%%%%%%%%%%%%%%%%
\section*{Appendix}
\setcounter{equation}{0}
\def\theequation{A\arabic{equation}}

In this appendix, we derive the effects of a finite $\eta$ coverage on
the extracted polarization parameters $p_H$ and $\rho_{00}$ for the
simple case where the parent hadrons are at rest. We assume that the
angular distributions of the decays, as shown in
Eqs.(\ref{Equation:pH0}) and (\ref{Equation:rho00}), are both uniform
in $\phi^{\star}$, the azimuth of a daughter particle. 

When a hadron decays into two daughter particles, let $\theta^{\star}$ be 
the angle between the global angular momentum $\textbf{L}$ and the
momentum of one daughter particle, we then have
\begin{eqnarray}
\cos \theta^{\star}=\cos (\phi^{\star}_{p} - \phi_{\textbf{L}}) \sin \theta^{\star}_{p}.
\end{eqnarray}
In the above, $\theta^{\star}_{p}$ and $\phi^{\star}_{p}$ are
respectively the polar angle and azimuth of the daughter particle in
the rest frame of the parent hadron.
When decay daughters can only be measured within a finite $\eta$ range
$|\eta| < \eta_{\rm max}$, the range of $\theta^{\star}_{p}$ is no
longer $[0, \pi]$ but is given by 
\begin{equation}
\theta^{\star}_{p} \in [\epsilon, \pi-\epsilon], {~\rm where~}
\epsilon = 2 {\rm ~atan} (e^{-\eta_{\rm max}}).
\end{equation}

To extract the polarization parameter $p_H$ from $\Lambda$ decays, one
can use either the average value of $\cos \theta^{\star}$ in
Eq.(\ref{Equation:pH1}) or $\cos (\phi^{\star}_{p} -
\phi_{\textbf{L}})$ in Eq.(\ref{Equation:pH2}) for the daughter protons.
However, a finite $\eta$ coverage gives the following results:
\begin{eqnarray}
\langle \cos \theta^{\star} \rangle
&=&\int \left (1+\alpha_{H}p_H\cos\theta^{\star} \right ) \cos
  \theta^{\star} \sin \theta^{\star}_{p}
  \; d\theta^{\star}_{p} d\phi^{\star}_{p} \nonumber \\
&& /\int \left (1+\alpha_{H}p_H\cos\theta^{\star}
  \right ) \sin \theta^{\star}_{p} \; d\theta^{\star}_{p}
   d\phi^{\star}_{p} \nonumber \\ 
&=&\alpha_{H}p_H \left [ 5- \cos (2 \epsilon) \right ]/12,
\end{eqnarray}
and
\begin{eqnarray}
\langle \cos (\phi^{\star}_{p} -\phi_{\textbf{L}}) \rangle
=\frac {\pi \alpha_{H} p_H}{8 \cos \epsilon} 
\left [ 1+ \frac {\sin (2 \epsilon)-2 \epsilon}{\pi} \right ].
\end{eqnarray}

Therefore, when one neglects the effect of a finite $\eta$ coverage
and still uses Eq.(\ref{Equation:pH1}), the following
polarization parameter will be extracted: 
\begin{eqnarray}
p^{\rm extracted}_{H}=p_H \left [ 5- \cos (2 \epsilon) \right ]/4,
\label{ph1extracted}
\end{eqnarray}
which is different from the real value $p_H$. 
We can see that the extracted value in this case is always no less
than the real value: it equals the real value when $\epsilon=0$ (full
$\eta$ coverage) but is 50\% higher than the real value in the limit 
of zero $\eta$ coverage. 
Likewise, if one neglects the effect of a finite $\eta$ coverage and still
uses Eq.(\ref{Equation:pH2}), the extracted polarization parameter will be
\begin{eqnarray}
p^{\rm extracted}_{H}=p_H \left [ 1+ \frac {\sin (2 \epsilon)-2
  \epsilon}{\pi} \right ] /\cos \epsilon.
\label{ph2extracted}
\end{eqnarray}
The above extracted value is also always no less
than the real value, and it is $4/\pi$ times the real value in the limit 
of zero $\eta$ coverage. 
The analytical results of Eqs.(\ref{ph1extracted}-\ref{ph2extracted})
are shown in Fig.\ref{fig5} and agree with the numerical
results from the Monte Carlo toy model.

For $\phi$ decays, one can use  the average value of $\cos^2 \theta^{\star}$ 
to extract the polarization parameter $\rho_{00}$. 
A finite $\eta$ coverage leads to the following:
\begin{eqnarray}
&&\langle \cos^2 \theta^{\star} \rangle
=1/80\times [133+401\rho_{00}+(4-172\rho_{00})\cos(2\epsilon) \nonumber \\
&&-9(1\!-\!3\rho_{00})\cos(4\epsilon)]/[7+3\rho_{00}+(1\!-\!3\rho_{00})\cos(2\epsilon)].
\end{eqnarray}
If there is no restriction on $\eta$, the above result reduces to 
$<\!\cos^2 \theta^{\star}\!>=(1+2\rho_{00})/5$.
However, if one still uses this relation 
when there is a finite $\eta$ coverage, 
then a different polarization parameter $\rho_{00}$
will be extracted. In the special case of no polarization
(i.e. $\rho_{00}=1/3$), we will have
\begin{eqnarray}
\rho^{\rm extracted}_{00}=\frac{1}{3}+\frac{5\sin^2\epsilon}{12}, 
\end{eqnarray}
which is shown in Fig.\ref{fig3} and agrees with the Monte Carlo toy
model results.

%----------------------------------------------------------------------------------------------------------------------------

\end{document}